\documentclass{kluwer}    % Specifies the document style.
\usepackage{graphicx}
\newdisplay{guess}{Conjecture}

\begin{document}                                                                                   
\begin{article}
\begin{opening}

\title{Galaxies at High Redshift}
\author{Garth \surname{Illingworth}}  
\runningauthor{Garth Illingworth}
\runningtitle{Galaxies at High Redshift}
\institute{UCO/Lick Observatory, Astronomy and Astrophysics Department,
University of California, Santa Cruz, CA \ \ 95064}
\date{October 15, 1999}

\begin{abstract}
Within just the last few years, we have advanced from knowing only a few galaxies at $z>2$ 
to having redshifts for nearly 1000 $z\sim $ 2--5 objects.  The majority of this sample 
has been detected through the photometric ``drop-out'' technique used so successfully 
by Steidel and his collaborators.  In addition, a handful of objects have 
already been detected at $z>5$, and we may already have a few objects at 
$z>6$!  These data, plus that at $z<2$ have resulted in a characterization of the 
star formation history of the universe, commonly known as the ``Madau plot'', which gives 
SFR per comoving volume {\it vs} redshift from the present day at $z=0$ to 
$z\sim 5$.  Recently,  we have realized that dust has significantly affected our
SFR estimates for the objects that have been detected in the optical, and 
have developed approaches to correct for that extinction.  We now also have increasingly good
evidence that a substantial fraction ($>$50\%?) of the high redshift star formation 
occurs in dust-enshrouded starbursts that were detected at 850 $\mu$m
with a submm bolometer SCUBA on the JCMT telescope.  These objects will typically be too faint to 
detect and measure redshifts in the optical, though those that can have already provided valuable
constraints.  The highest redshift objects at $z\sim 5$ and beyond are intriguing sources, but
test even the limits of Keck and of HST. The detailed study of such sources may well
only be practical for gravitationally-lensed objects, where the high magnification allows for
a much more detailed study of the proto-galaxy's structure, until even larger 
space-based (NGST) and ground-based (30-m plus) telescopes are developed.
\end{abstract}


\end{opening}

\section{Introduction}
 
Two new telescope/instrument combinations came on line in the mid 1990s
that resulted in a dramatic increase in our ability to observe distant
galaxies.  The refurbishment mission that added the WFPC2 camera to HST, 
along with its corrective optics, plus the commissioning of the 
Low Resolution Imaging Spectrograph on the Keck I telescope
provided new, and complementary, capabilities for observing faint, 
small, low surface brightness objects.   This led to a remarkable resurgence 
of interest in distant galaxies, and a large increase in the statistical
and quantitative veracity of the resulting data. 

HST WFPC2 imaging has provided us opportunities to resolve
distant galaxies, and extend our morphological and structural characterizations to
galaxies at much higher redshift -- the limitation now is largely one of S/N due to the 
effect of the cosmological $(1+z)^4$ surface brightness dimming and to the (often) large photometric
{\it K}-corrections
due to the rest frame shift into the UV.  
Some of the best examples of the imaging capabilities of HST have been seen 
in the multi-color images of the HDF and the ``Groth'' Survey strip in 1995, and the HDF-S in 1998,
and in the deep images of many intermediate redshift (i.e., $z\sim $ 0.2--1) clusters.

The multi-slit spectroscopy from Keck with the LRIS spectrograph has proven 
to be an ideal complement of the the HST imaging, allowing 30--40 galaxy redshifts to be measured 
at once to fainter than $I\sim 25$. With Keck plus LRIS it has become possible to routinely derive 
redshifts for objects several magnitudes fainter than was practical in
major (and important) surveys carried out earlier  this decade on 4-m class telescopes
(e.g., the CFRS, Lilly et al. 1995).  

The major developments of the last few years are:

(1.) ---  the measurement of large numbers of redshifts for 
$z\sim $ 3--4 galaxies by Steidel and collaborators;

(2.) ---  the identification of several strongly-lensed 
(and hence highly-magnified) $z\sim $ 2--5 galaxies;
   
(3.) ---  the detection of a number of $z>5$ galaxies;

(4.)  ---  the characterization of the star formation history of the universe by Madau 
and others (the ``Madau'' plot -- SFR {\it vs} redshift);

(5.)  ---  the recognition of the impact of dust on the UV fluxes from  high 
redshift galaxies;

(6.) ---  the discovery  of distant galaxies that are 
strong submm sources with the JCMT SCUBA submm detector.

(7.) ---  the utilization of increasingly  more accurate photometric redshifts;

The improvements that are planned for HST imaging with the ACS (the new Advanced Camera), which
will provide a gain of 10$\times$ in the area-throughput product (the figure-of-merit 
most appropriate for imaging surveys),
and the dramatic increase in the numbers of 6.5-m to 10-m ground-based telescopes 
with their next generation
optical and near-IR multi-object spectrographs will lead to an even greater rate of 
progress in the high redshift galaxy field within just a few years.

\subsection{Issues/Questions}  

While progress on distant galaxies has been rapid, there are a number of outstanding
questions about the highest redshift objects (those at $z>2$) that are the center of
a number of debates.  Examples of these are (i) what is the effect of dust on the SFR derived 
from the rest frame UV SED and flux measurements of $z>2$ galaxies, (ii) what are the redshifts of the 
SCUBA submm sources, (iii) what differentiates the optically-detected starbursts from the 
highly-obscured, submm sources, (iv) what are the masses of the $z\sim$ 2--5 
objects, (v) when did the first major star-formation events take place (thus forming the first metals), 
(vi) what are the physical conditions in high redshift star-forming galaxies, (vii) 
by what processes, and on what timescales,  were the galaxies we see today assembled, and (viii) how do
those timescales compare to the timescales for when the bulk of the stars were made?

These questions, and the issues they raise, will be noted during the discussion of the results
that are presented in the main body of this paper, and are issues that are encompassed in
many discussions
of galaxy formation and evolution.  As an introduction to developments in high 
redshift galaxies, and to provide a context for much of the later discussion, a number of items
are highlighted first, including (i) timescales, (ii) the recent baryon census,  (iii) 
the photometric ``drop-out'' technique that has been used so successfully
to find high redshift galaxies, (iv) the representation of the star formation history of the universe
through what has become to be known as the ``Madau'' plot, and (v) the importance 
of the HST deep imaging fields. 

These are followed by more detailed discussions of the results on (i) $z\sim $ 2--5 galaxies, (ii)
the SCUBA results, (iii) the latest detections of the very youngest  $z>5$ galaxies, and (iv)
the importance of strongly-lensed sources.  Finally, a few comments are made about the future
capabilities that will provide the observational framework for the further 
development of this field, and an assessment of some of the key issues about distant galaxies.

\section{Timescales}

It is valuable when thinking about distant galaxies to be calibrated on the timescales
involved. Based on a reasonable set of numbers, i.e., an open cosmology with $H_0 = 65$  and $q_0=0.05$, 
$t_0 =15$ Gyr, with the lookback time at $z\sim 1$ is about 50\% of $t_0$,
while the lookback time to $z=3 $ is about 11 Gyr. Or, in a way that is better for thinking about
formation timescales, the time since recombination to $z=10$ is about 1.4 Gyr, to $z=5$ is about 2.6 Gyr
and to $z=3$ is about 3.9 Gyr.  The timescales grow somewhat longer with a lambda-universe,
with $\Omega_{\Lambda} =0.7$ and $\Omega_m =0.3$, and dramatically shorter for an Einstein-de 
Sitter $\Omega_m =1$ cosmology (giving a rather unrealistically
short timescale for galaxy development by $z\sim 5$ of $<$1 Gyr!).

\section{Baryon Census}

A valuable framework for thinking about galaxy development was provided recently 
by Fukugita, Hogan and Peebles (1998).  Taking a wide variety of data sources,
 they did a census of the baryons at both low and high redshift.
While there are significant uncertainties associated with these estimates,
they do provide a useful framework for constraining where the mass is in galaxies by type.
For example, the low redshift census of the baryons in stars in galaxies shows that
the dominant sink of baryons in galaxies integrated over all time $t_0$ is spheroids.   
Ellipticals and the bulges of disk
systems currently contain, in their stellar population, about 63\% of the baryonic mass that is 
in galaxies (where this baryonic component is considered to be that in stars, stellar evolution 
end-products, or gas in the galaxy).  All disks contain only about 21\%
of the baryonic mass, while the extremely numerous
late type, and, typically, lower luminosity, galaxy population only has 2\% of the baryonic 
mass in stars.  The latter mass fraction
is small even compared the gas that now remains in galaxies.
The  gas in galaxies comprises 15\% of the total baryonic mass (integrated over all types).  
Thus a characterization of the star formation history of the universe is largely
a description of the formation of spheroid populations, and secondarily, of disks, though
if the timescales are quite different, one could still have a period where the dominant
process is disk formation.

\section{The Photometric ``Drop-Out'' Technique}

The Lyman break photometric ``drop-out'' technique was first utilized for the 
detection of high redshift quasars, but was then 
much more finely tuned as a detection technique for high redshift galaxies
by Steidel and his collaborators in the mid-1990s.  The technique relies on the 
large break in the continuum flux from an object that occurs at the 912 \AA \ Lyman limit 
from neutral hydrogen absorption in the line-of-sight. 
Multi-band images of a field containing high redshift galaxies can be used to 
identify those objects that have very red colors as a result of the redshifted Lyman
limit falling between any two filters.
The technique is refined
by having multiple filters that also can detect the smaller break at L$\alpha$
(1216 \AA ), and the rather blue continuum longwards of L$\alpha$.

As first used extensively by  Steidel et al. (1996) the technique was used to detect 
$z\sim 3$ objects by their lack of flux in the U-band (hence the descriptor ``U-band drop-outs'').
The technique was often applied by using three or four filters and defining a region
in the two-color plane in which such ``drop-outs'' were most likely to occur.
Steidel and his collaborators used UGR ground-based images, while, for the HDF, the four band images
allowed the use of a plane that was essentially $(U-B)$ {\it vs} $(V-I)$.  Examples of the two-color
plane and the selection function are given also in Dickinson (1998), as is a 
very instructive visual representation of the ``drop-out''  technique for a galaxy at $z=3$.

This technique has proved to be remarkably useful for detecting $z>2$ galaxies with ground (and space)
imaging in the ``optical'' ($\sim$0.3--1 $\mu$m).  A key advantage of this technique is that 
it is essentially free
of selection effects, with little contamination from low redshift objects.
All high redshift objects above a given magnitude limit will be detected, provided the S/N is high 
enough in all the bands, particularly the bluest band where an upper limit must be established.
Occasionally red stellar objects and dusty galaxies contaminate the sample, but the fraction
is small with high S/N images.  

The only objects which might still be detectable spectroscopically, but would be missed
by this technique are those with a very strong emission line and very weak continuum fluxes 
(typically the strong line would be Ly$\alpha$).  Stockton (this volume) and Hu et al. (1999)
discuss examples of such objects, and also discuss narrow-band imaging searches for 
detecting such objects (see also \S 8 below).

\section{The ``Madau plot'' -- SFR {\it} vs redshift}
 
While a number of authors over the years have discussed
the star formation history of the universe, one of the clearest
discussions of this issue has been provided by Madau (see, e.g., Madau et al. 1996; 
Madau 1997; Madau, Pozzetti and Dickinson 1998).
The form that was used by Madau (1997) is shown in Figure 1, and gives the 
star formation rate per comoving volume  {\it vs} redshift.  This was one of the early
papers from Madau and shows a rise from $z\sim 5$ to $z\sim $ 1--2, and then a comparable 
drop to the present day. This form of the SFR {\it vs} redshift figure has become
known as the ``Madau plot''. If plotted  versus time, instead of redshift, the star formation
history is more symmetric with a clear peak around a lookback time of $\sim 60$\% of $t_0$.

\begin{figure}
\centerline{\includegraphics[width=18pc]{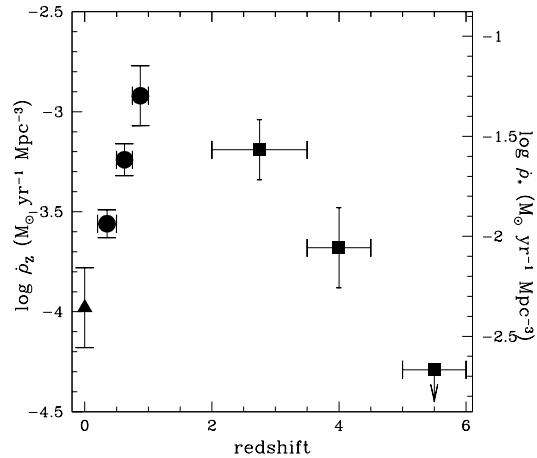}}
\caption{The star (and element) formation history of the universe, from the HDF
(Madau 1997). The early data used for this figure indicated a strong peak in 
the SFR at about $z\sim $ 1--2. See below for the form of the ``Madau plot''
just 2 years later.}
\end{figure}

However, a number of developments (e.g., dust corrections, submm results, new estimates of the SFR
at low redshift) are changing the shape of the SFR {\it vs} time, and making the variation 
since $z\sim 5$ much less. For example, one of the most significant developments of 
the last few years has been the recognition that dust extinction is skewing our view of 
distant galaxies.  In particular,  the rest frame UV SEDs (spectral energy distributions) that are 
measured from optical and near-IR observations invariably have slopes indicative
of modest reddening.  Since even small amounts of reddening significantly affect the 
the UV fluxes that we observe the resulting corrections can be large (a factor of 2 or more).
This important development will be discussed further below.

\section{HST WFPC2 Imaging}

The impact of HST on the field of galaxy evolution cannot be overstated.
The revolution that we have witnessed in the last five years in the field of distant galaxies
is due primarily to the high resolution imaging capability of the WFPC2 camera (though Keck 
spectroscopy has also played a key role).
The morphological and structural information that can now be obtained for 
high redshift galaxies has placed the studies of distant galaxies
onto a much more quantitative footing.  So much so, that it is hard to conceive of this
field without the low background, high resolution, wide-field imaging capability  that we get from HST.
The impact that the Hubble Deep Field (Williams et al. 1996) had on this field is 
well known.  The number of papers generated from this tiny, five (arcmin)$^2$ region
is remarkable.  Even now, four years after the data were taken, it is still
a joy to look at the color images of the HDF.  The range in scales and morphologies
of the galaxies in this image is astonishing, as can be seen in Figure 2 (See 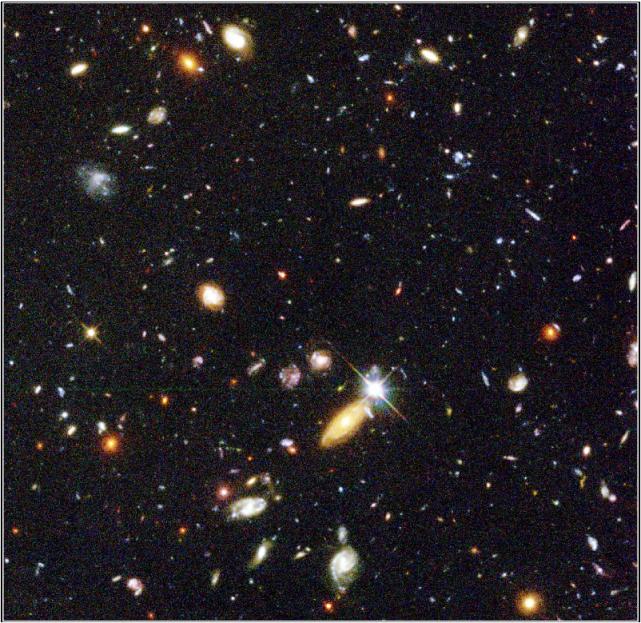),
where an expanded  section of the HDF is shown.

The HDF-S added a comparable region to the southern sky opening up more opportunities for 
ground-based imaging (particularly JHK IR imaging) and spectroscopy (e.g., with the VLT telescopes).
A number of additional fields have been imaged with HST, one of which is proving
to be valuable in a quite different way.  Ground-based multiobject spectrographs typically have much larger 
fields than the WFPC2 field, and so single HST pointings are not ideal for efficient
spectroscopic followup on the ground.  For example, the LRIS field on Keck covers about three 
contiguous WFPC2 areas, and the next generation of even wider field,  multiobject spectrographs
will cover more than twice the field of LRIS (e.g., DEIMOS on Keck has over 15$'$ of useful field 
in one direction) and so 6-7 contiguous pointings in a row are really needed from HST.
Thus, HST imaging surveys like the WFPC2 GTO team survey strip that has  28 contiguous WFPC2
images (2.5$'\times$42$'$) in F606W (2800 s) and F814W (4200 s) are optimal for the new 
redshift surveys, like the Keck DEEP project (Davis and Faber 1999; Koo 1999).  
While not as deep as the HDF, the ``Groth strip'', as the GTO team 
survey area is known, is proving to be an invaluable survey region.  The magnitude limit
for obtaining photometric and structural parameters for distant galaxies is comparable
($\sim$23 mag) to that for which redshifts can easily be determined using the Keck telescope
and a multiobject spectrograph.  

The future for HST imaging, particularly with the new wider-field, more efficient ACS, 
will likely be a combination of very deep, multiband images (like the HDF) for photometric 
redshift studies of the very faintest objects, and larger field, multiple-pointing,
multiband imaging surveys (like the ``Groth strip'') that can be combined with the 
redshifts and velocity widths determined from spectroscopic surveys with
the many 6.5--10 m ground-based telescopes.

\section{Galaxies at $z\sim 3$}

One of the most important developments of the last decade has been the 
systematic study of $z\sim3$ galaxies carried out be Steidel and his 
co-workers (see e.g., Steidel et al. 1996).  Using largely ground-based UGR photometry 
to find $z\sim 3$ candidates from U-band ``drop-outs'' Steidel and his collaborators 
have obtained redshifts with Keck and confirmed some 700 $2.5 < z < 3.5$ galaxies.
This is a remarkable accomplishment, when one considers that only five
5 years ago the number of known galaxies  at $z>2$ was but a handful.

By extending the photometry to the red, with GRI images, Steidel et al. (1999) have obtained a sample
of ``blue drop-outs'' (strictly G-band drop-outs, but usually referred to as ``blue'' as opposed to 
``uv'' dropouts) that have already yielded a large number of $z\sim 4$ galaxies. The redshift-confirmed
sample of $3.5<4<4.5$ galaxies is already nearly 50 objects (Steidel et al. 1999).

\addtocounter{figure}{1}
\begin{figure}
\centerline{\rotatebox{-90}{\includegraphics[width=18pc]{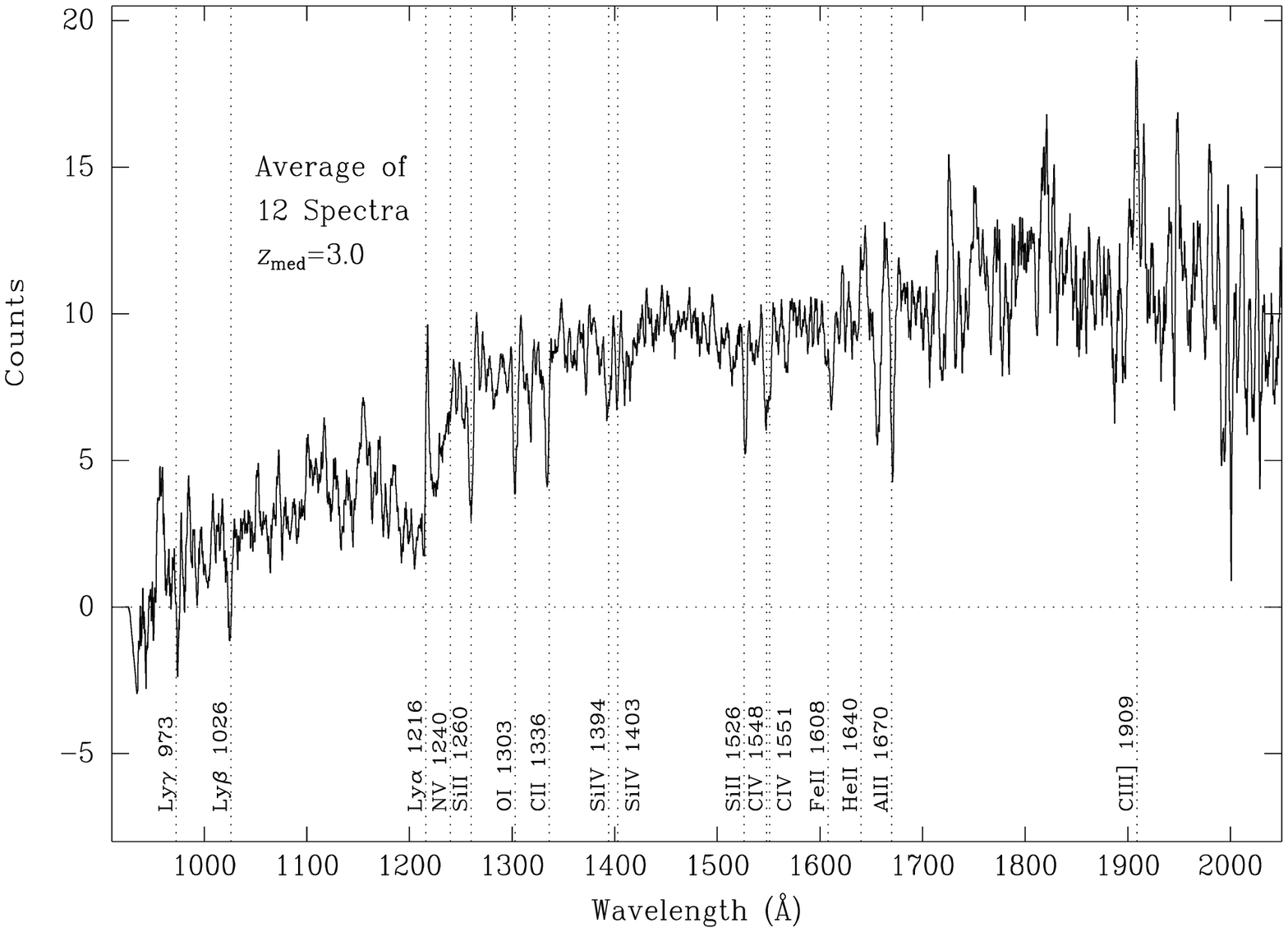}}}
\caption{The composite Keck LRIS spectrum of the Lowenthal $z\sim 3$ galaxies in the HDF. 
Note the weak Ly$\alpha$ emission, the damped absorption at Ly$\alpha$, the ISM and stellar
wind absorption lines, and the UV continuum slope.}
\end{figure}

When HST data is available, the $z= $ 3--4 galaxies can be analyzed for structural characteristics 
and length scales.   A study that showed the value of combining 
redshifts with HST imaging was that of Lowenthal et al. (1997) for a small
sample of $z\sim 3$ galaxies in the HDF.  The composite spectrum of the galaxies 
that were measured with Keck is shown in Figure 3.  The spectral characteristics are
very similar to low redshift starbursting systems (typical SFR in this HDF sample are 
around 10 $M\odot$ yr$^{-1}$).   The characteristic weak Ly$\alpha$,
as well as the many absorption lines from the ISM and stellar winds can be seen in this composite spectrum.
The galaxies that have been detected in this sample (and in the Steidel et al. survey) typically 
have luminosities $L^*$ or greater, and have quite small half-light radii ($\sim 0.3''$).
Such measurements allow us to compare these objects with low redshift galaxies.
As can be seen in Figure 4 (from Lowenthal et al. 1997),
they are typically more luminous than low redshift galaxies of comparable 
size, but it is instructive to note that the stellar population from such a young burst
will dim by some {\it five} magnitudes, and so these objects could well be the precursors of 
local sub-$L^*$ galaxies (of the bulges?).  However, much can, and likely will happen to these objects 
in the 10 Gyr between $z\sim 3$ and now  and so it is not obvious what such objects will become in individual
cases --   though since most stars will end up residing in bulges, as was noted above from the 
results of Fukugita, Hogan and Peebles (1998),  it is almost a no-brainer to
note that these are likely, on average, to be the precursors of present day bulges.

\begin{figure}
\centerline{\includegraphics[width=18pc]{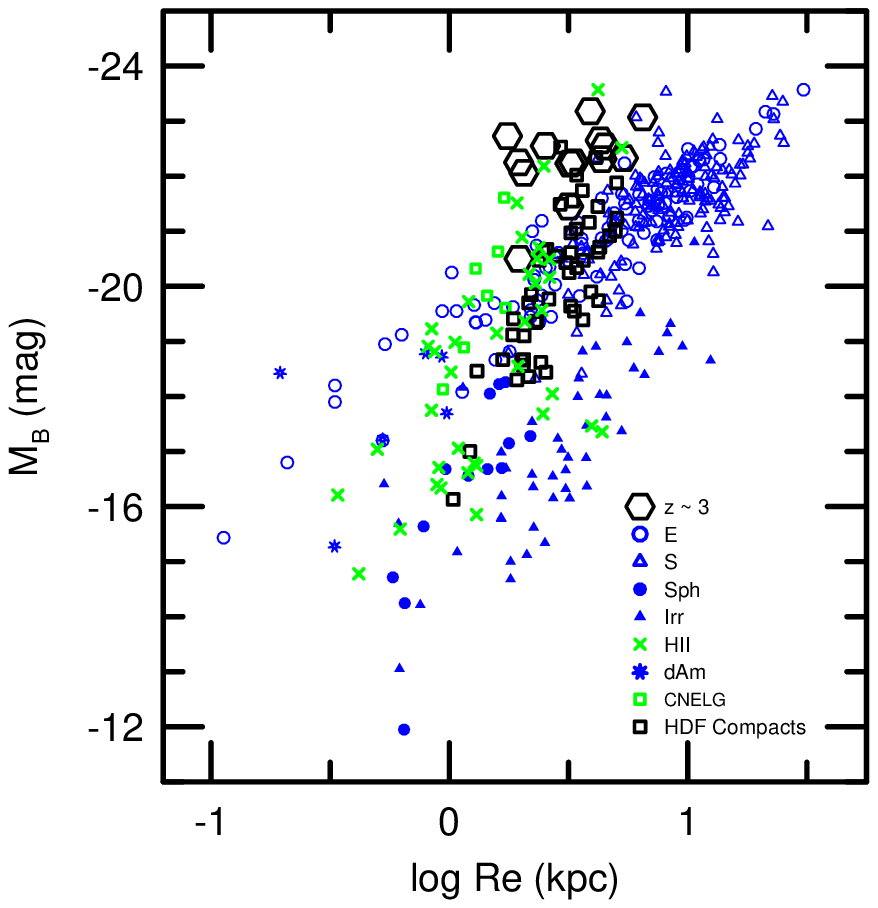}}
\caption{The absolute magnitude {\it vs} half-light radius for a large sample of nearby galaxies
of many morphological types, overplotted with the HDF $z\sim 3$ sample (from Lowenthal et al. 1996).
The $z\sim 3$ galaxies are luminous, but quite compact.}
\end{figure}

One other conclusion that can be drawn from the composite spectrum of Figure 3
is that the typical $z\sim 3$ starburst is extincted. The slope of the UV continuum 
around 1500 \AA \ can be used to estimate the level of extinction. It turns out, as 
might be expected, that the extinction correction is an important one, with 
$E(B-V)$ typically being found to be  $\sim$0.3 (Steidel et al. 1999).  The resulting 
correction for the star formation rate does depend sensitively, however,  on the form of the dust
extinction law that is used. For example, Dickinson (1998) shows that the ratio
of the mean corrected SFR to the uncorrected SFR for $z\sim 3$ galaxies
was a factor 7 (!) for a Calzetti reddening law (Calzetti, Kinney and Storchi-Bergmann 1994), 
but only a factor 2 for an SMC 
reddening law, with the Calzetti law giving much greater dispersion due to its rather grey
behavior.  Since the Calzetti reddening law was derived for (low redshift) starburst galaxies 
it might seem more appropriate to use it for the high redshift starbursts, but metallicities,
and the (possibly) different environment may lead one to a different reddening law --
but it will be difficult to verify what the different law should be.

The net effect of both the larger sample of galaxies at high redshift and the 
corrections for extinction has been to raise significantly the star formation rate at 
$z>2$ where the ``drop-out'' sample have been the dominant source of constraints on the SFR
{\it vs} redshift. With no extinction corrections the rates have increased with the larger samples so that
there is only a modest factor of 2 increase from the SFR at $z\sim 4$ to the peak SFR at $z\sim 1.5$,
in contrast to the original factor of nearly an order-of-magnitude from the HDF data alone
(small number statistics?).  Correcting for extinction, based on a Calzetti reddening law,
leads to a further increase of the SFR at redshifts $z>2$ to where the SFR($z$) relation
is essentially unchanged at higher redshifts, from redshift $z\sim $ 4--5 to $z\sim 1.5$.
The ``Madau plots'' incorporating these new data are shown in Figure 5 (from Steidel et al. 1999).

\begin{figure}
\centerline{\includegraphics[width=18pc]{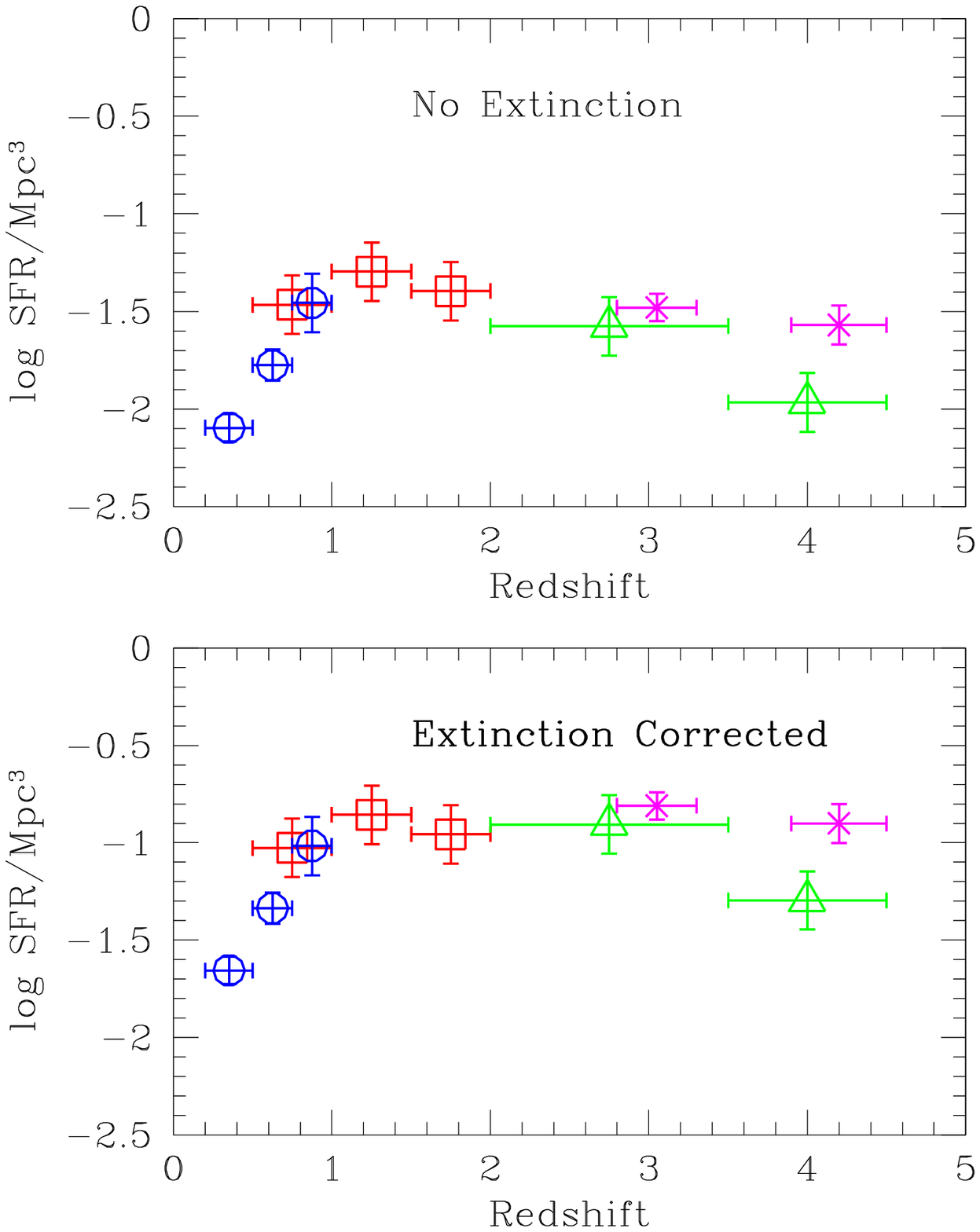}}
\caption{The recent derivation by Steidel et al. (1999) of the ``Madau plot''
showing the effect of using their much larger sample, and of the importance of 
the corrections for extinction.  The HDF (lower symbols at $z=3$ and $z=4$) 
contained fewer high-redshift galaxies than 
the typical fields of Steidel et al., leading to an underestimate of the SFR at $z\sim $ 3--4
(as in Figure 1).}  
\end{figure}

\subsection{Masses}

One of the most important observational questions for these high redshift objects 
is ``what are their masses?''.  Masses are a crucial aspect of relating the objects 
we see at high redshift to those at low redshift, and for comparisons with the 
predictions of theoretical models.  However, the strongly star-forming, compact regions that 
dominate in what we observe in the high redshift galaxies makes it very difficult to
derive the masses of the $z\sim 3$ objects with any confidence.  For example, strong 
outflows are seen in such objects (see, e.g., Franx et al. 1997), and so measurements of
velocity widths $\sigma$ from absorption, or emission lines, are likely to lead
to overestimates of the gravitational masses. Yet other effects, namely dust, the small sizes
of the star forming regions, etc., can lead to underestimates of the mass.  It is not clear, 
given the limited data that we can get, even what the sense is of the overall correction,
and so it has not been possible to derive reliable mass estimates from the observed 
velocities and length scales. 

Pettini et al. (1998) and Lowenthal et al. (1998)
are attempting to derive constraints on the masses,
by looking at optical nebula lines, redshifted into the 2 $\mu$m region, or by using 
component velocity differences, but the results are still limited and very uncertain.

\begin{figure}
\centerline{\includegraphics[width=18pc]{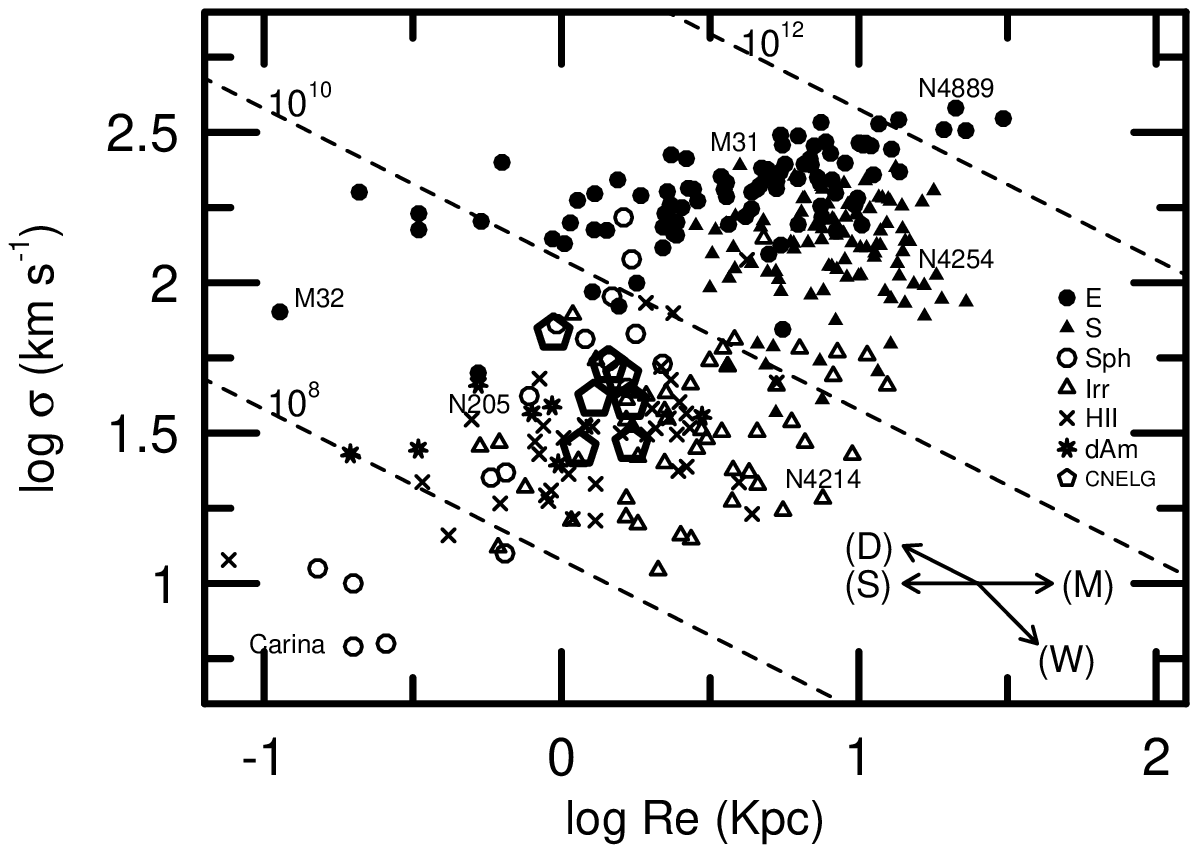}}
\caption{Half-light radius {\it vs} velocity width for nearby and low redshift
galaxies, from Guzman et al. (1996). The corresponding mass scale are indicated by the dotted lines.
The effects of various physical processes are indicated also in the lower right -- see Guzman 
et al. for details.  It is clear that a wealth of information can be derived for high redshift
galaxies once they can be placed in this quantitative plane.} 
\end{figure}
 
The value of deriving velocity and length scales in a way that is consistent with that 
done for low redshift galaxies is apparent from Figure 6.  This figure is from
Guzman et al. (1996), and compares intermediate-redshift, compact, 
star-forming galaxies with low redshift galaxies,
to establish their likely mass scales.  Such a comparison would be very valuable to have also
for high redshift galaxies.

\subsection{Dust, the UV and the Submm}

As noted above the issue of dust and the extinction of the optically-detected, UV rest-frame, 
sources has been the subject of considerable discussion over the last few years.
In parallel with efforts to characterize the impact of dust on optically-detected 
high redshift galaxies, a very dramatic
set of results has become available over the last year on sources at high redshift
that are emitting predominantly in the submm.  These detections were made at 850 $\mu$m with a new
bolometer SCUBA on the 15-m JCMT telescope in Hawaii.  It's large  15$''$ `beam', combined with the 
generally low UV and optical fluxes from these highly extincted sources, has made it hard to 
reliably identify the corresponding optical source and derive a redshift.  
A novel approach at improving the 
positions, by using radio continuum flux, has improved the source 
``identification'' rate, though this may only be useful for the most luminous sources.   
 Nonetheless, enough redshifts have been 
determined, which, when combined with indirect arguments, suggest that such objects could well
be contributing as much (UV-equivalent) flux at high redshift as the optically-detected sources, 
if not more.
Whether this is entirely due to starbursts, or whether AGN activity is also a contributing factor
remains to be determined. However, it is likely that they are at least comparable to the 
optically-detected sources in their contribution to the overall SFR at high redshift, and may 
actually be the dominant population in the ``Madau plot'' at $z>1$!

Such sources will be readily detectable with ALMA, and observable over a wide redshift range
because of a shape of their SED.  These sources have been detected and measured at 850 $\mu$m,
which lies on the long wavelength side of the black-body distribution, 
such that as the redshift increases the flux increases (unlike the usual situation in the optical
where the K-corrections are negative), and the increase is such as to largely cancel out the 
cosmological $(1+z)^4$ surface brightness dimming (see Sanders, this volume, 
for a good explanatory figure).  Much remains to be learnt about these 
important new objects.

\section{$z>5$ Galaxies}

Just three years ago, the first galaxy was found that had a higher redshift
than the then highest redshift QSO; such an event was expected given that galaxies
presumably predate QSOs, but this was the first time since the 
discovery of QSOs in the 1960s that this had happened.  This object was at $z=4.92$ (Franx et al. 1997).
It identified $z > 5$ as the time when we might begin to see the development of substantial 
baryonic potential wells.  Since then,  the highest redshift QSO has crept over $z=5$, 
but the highest redshift galaxy has jumped to at least $z=5.74$, and possibly even to $z=6.68$.

The three best determined $z > 5$ objects are at $z=5.34$ (Dey et al. 1998),
$z=5.60$ (Weymann et al. 1998), and $z=5.74$ (Hu et al. 1999).  All these redshifts were
measured from Keck LRIS spectra, and all show Ly$\alpha$, though all are faint with integrations 
ranging from 4 to 10 hours, as  required to get adequate S/N in the emission line.  
The continuum fluxes are low, since these sources typically have $I_{AB}$ magnitudes around 26-27, 
and hard to detect against the bright night sky (Ly$\alpha$ falls at 800 nm at 
$z=5.6$).  The detections are almost certainly real, since the objects are measurable 
in deep imaging data, and the expected flux decrements due to the continuum breaks
at Ly$\alpha$ 1216 \AA \ and the Lyman limit at 912 \AA \ have been seen.  In addition, the line profiles
display the expected asymmetry due to blue edge absorption in Ly$\alpha$ in the outflowing ISM 
associated with the starburst (see e.g., Franx et al. 1997).  
The highest redshift object, at $z=6.68$ (Chen, Lanzetta and Pascarelle 1999), is a less certain 
detection, given that it is faint and the data are relatively low S/N.  

It is clear that we are pushing to $z\sim 6$ and beyond, but it is also clear that 
further progress really requires good near-IR imagers and spectrographs.  At $z=6.68$, 
Ly$\alpha$ falls at 940 nm, where typical CCDs have quite low quantum efficiency.
Another interesting issue with these high redshift objects is what will we actually do with 
them?  They are so faint that it will be extremely difficult to to obtain high S/N data, 
even in images.  As noted, they are fainter than $I_{AB}=26$, with typical line fluxes 
that are very low, and, even with Keck, integrations of 4-10 hours provide little
more than redshift detections.  If we are to get more detailed information about
such high redshift objects then a different approach must be used.

\section{Strongly-Lensed Galaxies at High Redshift}

Over the past few years a number of strongly-lensed distant galaxies have been found where
the magnification is high enough that the source object can be studied at a level of detail 
that is impractical for typical galaxies at that redshift.  An excellent recent example is
the very good paper by Pettini et al. (1999) on the $z=2.73$ strongly-lensed object MS 1512-cB58.   
Another example is the strongly-lensed object CL 1358+62\_G1 in the $z=0.33$ cluster 
CL 1358+62 (Franx et al. 1997).  This object is at $z=4.92$.  The large magnification 
($\sim$10$\times$) makes
this a particularly important pathfinder for assessing the structure of galaxies at high redshifts.
While it would be valuable to have other examples to enlarge the sample, this remains the best known 
at such a high redshift.

The cluster and its arc are shown in Figure 7 (See 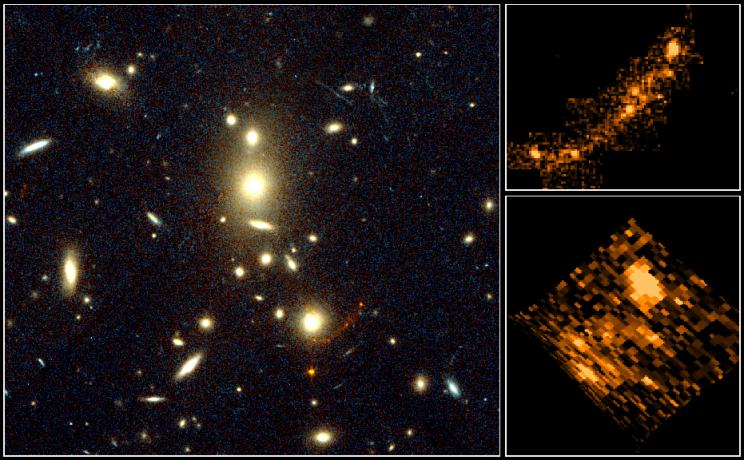), along with the arc prior to reconstruction,
and after reconstruction into the source plane.  The resolution in the reconstructed image is 
better than 20 milliarcsec, comparable to what one will get with adaptive optical systems on
8-10 m telescopes in the near-IR, though the sensitivity to low surface brightness extended structures
will be much less in such ground-based data.  This resolution corresponds to about
200 pc.    

The object in the reconstructed image covers about 7 kpc, with several star forming knots 
interspersed throughout the galaxy.  The knots dominate in that they contribute about 75\%
of the  flux in the object, with the brightest knot alone contributing about half the flux.
This corresponds to more than $10^{11} L_\odot$ in a single region with a half-light radius
$r_{1\over 2} \sim 20$ mas or a FWHM of about 300-400 pc.
The SFR implied by the observed UV flux, under the assumption of no extinction,  is about 50
$M_\odot$ yr$^{-1}$.  However, it is clear from the combined HST optical and Keck near-IR 
data (Soifer et al. 1998) 
that the source is reddened with $E(B-V) \sim 0.3$, indicating that the SFR needs to be increased
significantly to $>10^2 M_\odot$ yr$^{-1}$, possibly $>>10^2 M_\odot$ yr$^{-1}$, 
depending on  the reddening law and the dust distribution.  
The value of the IR data can be seen in Soifer et al. (1998) in the fits of the Bruzual and Charlot
models, with Calzetti extinctions, to the measured HST WFPC2 F606W and  F814W fluxes, and the Keck
NIRC near-IR J, H, and K fluxes.  Both the instantaneous burst and continuous star formation
models need extinctions $E(B-V) \sim 0.3$ to match the observed fluxes, given the constraint that the 
burst must contain stars less than $10^7$ yr in age so as to ensure the UV flux required for
the strong Ly$\alpha$.

The knot mass can be estimated from the IR luminosity, assuming a Salpeter IMF, 
and is $\sim 5\times 10^9 M_\odot$.  Given the scale size of the knot, this implies a
velocity dispersion of $\sigma \sim 200 $ km s$^{-1}$.  The size of the region, and such a velocity
dispersion, is typical of many present day bulges.

Another interesting aspect of this object is that there are multiple star-forming knots within 
a more extended structure.  This morphology is characteristic of the ``christmas tree'' model
of star-formation in distant galaxies where different knots may ``turn on'' at different 
times, as was discussed by Lowenthal et al. (1997).  A more detailed analysis of the implications for star
formation at high redshift is to be found in Illingworth et al. (2000).

\section{Future Capabilities}

While progress on high redshift galaxies has been quite dramatic over the last few years
a number of changes are underway that will provide a similar quantum step in capability
in just the next 4-5 years.  The new facilities and instruments can be put into three groups.

{\it Large Ground-Based Telescopes:} ---   
there will be roughly an order-of-magnitude increase in the number of large ground-based 
telescopes, from the two Keck 10-m telescopes, as the four ESO VLT 8-m telescopes 
are commissioned and a further 8--10 large ($\ge 6.5$ m) O/IR telescopes become 
operational (the $\sim $16 are evenly split N and S).  These will have efficient optical and 
near--IR ($\sim $ 1--2 $\mu$m) imagers,  and multiobject spectrographs, such as DEIMOS (for Keck)
and NIRMOS and VIRMOS (for the VLT), giving large throughput gains.
The potential of adaptive optics for studying distant galaxies is substantial,
and is part of the focus of a major new adaptive optics center funded by the 
NSF at the University of California, Santa Cruz  -- the Center for Adaptive Optics (CfAO).

{\it Hubble Space Telescope:} ---   there will be substantial increase in HST's capabilities, through the 
development of new imagers -- the typical gains over current instrumentation
are $>$10$\times$.  These instruments include the wide-field CCD  Advanced Camera 
with its SDSS filter set (as well as a large complement of narrow--band and other 
broad--band filters).  The ACS will be launched on the HST servicing mission SM-3B in 2001. 
The  WFC3/IR UV-IR imager that adds wide-field IR imaging capability in J and H to HST
is slated for launch on the HST servicing mission SM-4 in 2004.   

{\it Submm/IR:} --- there will also be major developments in submm imaging, starting 
with upgraded  SCUBA-like imagers on the JCMT, and followed by new 
interferometer arrays, particularly ALMA, the Atacama Large Millimeter Array. 
ALMA will be an extremely powerful facility for studying dusty objects over a 
large range of redshifts. Ultimately, NGST should allow us to directly image 
extremely young protogalaxies during their first major star forming events.

The key to characterizing the universe at intermediate ($z\sim 1$) and high ($z\sim $ 3--5+) 
redshifts, and beyond, will be 
through extensive surveys to very weak flux levels.  Such combined redshift and imaging surveys
(e.g., DEEP -- see Davis and Faber 1999; Koo 1999) should produce samples with a level
of statistical robustness that will be unheralded for distant galaxies.

\section{Summary}

As with any subject in such an infant and  dynamic state there seem to be more questions than 
answers.  For example, some key ones that intrigue me with regard to high redshift galaxies
are:

(1.) --- What are the physical conditions in $z\sim $ 3--4 objects, and in particular, 
what are their mass scale lengths, masses, and metallicities?  In addition, what is the nature,
location and extent of the star-forming regions in these galaxies, and what are their 
merging rates? What is the distribution of the dust, and can we characterize the dynamical
conditions and the extent of the outflows in the ISM that envelopes their star-forming regions?

(2.) --- What are the properties of the youngest objects that we see, those at $z>5$?
Again, what are their metallicities, their mass scale lengths and their masses, and what
is their space density and luminosity function?  The strong Ly$\alpha$ lines, 
relative to the continuum, in contrast to the $z\sim $ 3--4  galaxies suggests that they are less
dusty, but is this a selection effect?  How much dust do ``typical'' $z>5$ objects contain?

(3.) --- What is the integrated SFR from $z\sim 5$ to $z\sim 1$ in the dust-enshrouded
(submm) sources?  How does it compare to the integrated SFR seen in the optical-UV 
(when corrections have been applied for the more
modest dust-absorption typically found in these optically-detected sources)?

(4.) --- Are we missing a whole set of objects with higher reddening/extinction that the 
typical $E(B-V) \sim 0.2-0.3$ sources found in the $z\sim $ 3--4 ``drop-out'' sample, but which would not 
contain enough dust to be detectable at 850 $\mu$m with SCUBA?

(5.) --- What is the form of the extinction at high redshift -- and can it be characterized
by a single  reddening law?

These questions notwithstanding, we have made substantial progress over the last 
few years.  Given that few galaxies were known at redshifts $z>2$ just five years ago, it is 
remarkable that we now can identify the time evolution of the star-formation
rate in galaxies from redshifts $z\sim 5$ through to the present day.  Many uncertainties remain, 
particularly with regard to the overall contribution from the dust-enshrouded submm sources,
but the most likely situation is that by $z\sim 5$, within 1-2 Gyr of recombination, 
the SFR per comoving volume element had reached a level comparable to that at all redshifts 
down to $z\sim 1$, after which it decreased somewhat to the present day.

Interestingly, very simple arguments based on the census of the baryons at the present day, 
the ages of the stellar populations in ellipticals (see e.g., van Dokkum et al. 1998),
plus the sizes and densities of the star forming regions at $z\sim 2-5$ suggest 
that most of the star formation we see at $z>2$ is making bulges.

\acknowledgements

I would like to acknowledge many valuable discussions with numerous colleagues --
the remarkable progress over the last few years is due to the imagination
and energy of many in our small community who have worked so hard to bring 
us state of the art facilities that open up new horizons.
Their efforts, and those who use them imaginatively to further our understanding, and those who
model the results and provide the theoretical underpinnings together make it 
a delight to be part of this field.  I am also particularly grateful to those 
in our funding agencies and national centers who work, often under great pressure, 
to bring us the facilities and funding that allows our community
to reach far and to do it so rapidly.  Last but most importantly,
this was a timely and fascinating conference and I am very grateful to our
energetic organizer, David Block, for an excellent conference,
and to Margi Crookes for her dedication to making everything work well, 
and to all those who helped them.  The  financial support from the 
Anglo-American Chairman's Fund and SASOL was essential to making this 
such a successful conference, and I would particularly like 
to thank those organizations.

\end{article}
\end{document}